# Fast and reasonable Installation, Experience and Acceptance of a Remote Control Room

R. Eisberg[1], E. Garutti[1], R. Kammering[1], A. Kaplan[1], S. Karstensen[1],
B. Lutz[1], N. Meyer[1], R. Pöschl[2], B. Warmbein[1]

1 – DESY Hamburg – Notkestrasse 85, 22607 Hamburg– Germany

2 – LAL– IN2P3, B.P. 34, Bat. 208 Orsay – France

Remote control systems are becoming more and more important to give us the flexibility to control facilities, provide assistance and intervene in case of problems at any time and from every place.
As a global operating group CALICE [2] with approx. 220 members worldwide is dependent on using a remote control system for shifts and monitoring of the data taking. CALICE has at present installed its detector at Fermilab, Chicago, where will run test beam experiments for the next year. The components of the remote control system and kind of use are presented here.

## 1 Main Criteria

One very important issue in the decision of installing a remote control facility is also the price to work-quality ratio. To install a remote control room should have a negligible impact on the overall project cost. This project has been realized within an adequate time, a minimum amount of manpower and maintenance but with a maximum efficiency.

Following items must been integrated:
- Web Based (no special software needed)
- Easy and fast to implement (<4 weeks) and easy to maintain
- Not too expensive (<10000 €) and nice to use (just start everywhere)

The realization of a remote control system considering the above mentioned items are shown here.

## 2 Components

The components of the remote control structure cover a web based secure global desktop, which allows every user to secure single sign on login, if necessary kerberized, to the control room operating systems, without having detailed knowledge of the operating systems themselves.
Additionally a conference system with a permanent connection between remote and on-site shifts is implemented to give all participants the possibility to work "in the same room".
Last but not least a camera system with possibility of remote control options has been installed. This system consists of several cameras, which allow fast, easy and very accurate observation of all experimental areas. This should give experts the opportunity to diagnose problematic situations without direct access to the experimental enclosure, and provide feedback to the crew in situ.

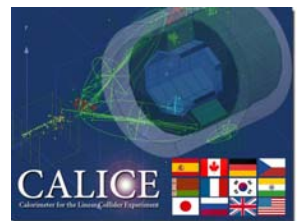

**Figure 1:** CALICE logo

## 3 CALICE at FERMILAB

The Control structure at Fermilab is divided into three parts:
- Control system for controlling and monitoring
- Conferencing for getting personal contact and
- Observation of the technical area.

### 3.1 Control System Fermilab side

The control system contains all parts to drive and steer the CALICE experiment, which subsisted out of the three detector parts ECAL, HCAL and Tailcatcher. Each system has its own slow control but all systems are conflated within the DAQ system.
For easy access and redundancy, two portals are included in an array, both equipped with a sun secure global desktop [5] web server and single sign on identification system, which is also connected to the kerberized Fermilab network. The entry point is stretched over two Linux PC's because of redundancy reason. Both PC's are connected within an array. That will give the administrator the possibility to interact on only one of the two and the database will be automatically updated on all SSGD PC's within this array.

### 3.2 Conferencing System

A Polycom conferencing system is installed in the CALICE control room at FNAL to have an easy possibility to communicate. We took the provided hardware by Fermilab, because of their good experiences with this kind of solution. The hardware is just connected to the Ethernet and configured in that way, to make a connection to the ESnet (see chapter 5 for more information). The Polycom hardware provides stereo microphones, echo canceling, remote configuration possibilities and connection for external monitor devices.

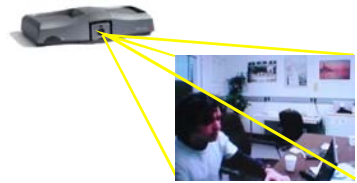

**Figure 2:** Polycom System

### 3.3 Observation System

To observe areas and give assistance in technical cases of doubt, web based remote controlled cameras are used. We decided to use fully web based controlled cameras to prevent our system from additional hardware and hence from additional installation and maintenance work. Remote software updating is also possible as remote configuration. The cameras can be connected either by a common web browser or special software with the possibility of alarm features and/or saving the images as movies.

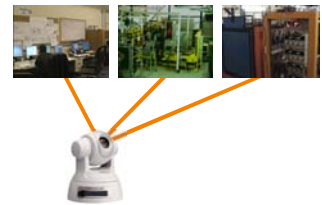

**Figure 3:** Webcam

## 4 Control Room DESY side

The control room is equipped inside a special designed room for the purpose to get a nice and quiet surrounding for the remote shift crew.

### 4.1 Control Console

Two DESY standard computers on the client site, each with 4 Monitors and a Java enabled browser - that's all.

The idea behind this simple structure is also simple: we integrated all intelligence and communication hard and software on the Fermilab site. No special knowledge is necessary, apart from how to install a computer and use it, to make a connection to the CALICE control system at Fermilab.
This does also mean if another person or institute want to get control of the CALICE experiment, a single https connection is necessary.

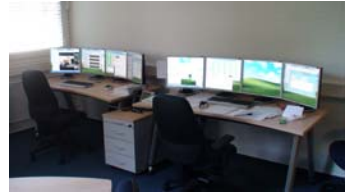

**Figure 4:** Control Consoles

### 4.2 Conference System

The counterpart of the Fermilab conference system comprise out of a wide TV screen, one camera, a echo cancellation audio table system and a computer system including the software, again from Polycom. An easy user interface helps the shift crew to get fast connection to the Fermilab control room in case of a non existing connection. The feeling is like that the conference participants are just looking through a window.

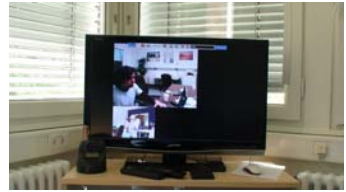

**Figure 5**: Conference System at DESY

### 4.3 Control Room

To give the shift crew the possibility to do good work, a pleasant area has been created. The room was equipped with well formed furniture, blue carpets, chairs, a proper blue wall as well as a sliding door which gives the feeling of expanse. Noise cancellation of the outer corridor helps the working personal to be concentrated on their work.

## 5 Conference System technique

The ESnet is the Conference Host of CALICE. ESnet, the Energy Science Network is a high-speed network serving thousands of Department of Energy scientists and collaborators worldwide. A pioneer in providing high-bandwidth, reliable connections, ESnet enables researchers at national laboratories, universities and other institutions to communicate with each other using the collaborative capabilities needed to address some of the world's most important scientific challenges.

Managed and operated by the ESnet staff at Lawrence Berkeley National Laboratory, ESnet provides direct connections to all major DOE sites with high performance speeds, as well as fast interconnections to more than 100 other networks. Funded principally by DOE's Office of Science, ESnet services allow scientists to make effective use of unique DOE research facilities and computing resources, independent of time and geographic location.

The same kind of network, but with the possibility of a higher bandwidth for HD cameras is existing in Germany, called DFN - Deutsches Forschungs Netzwerk

## 6 Electronic logbook

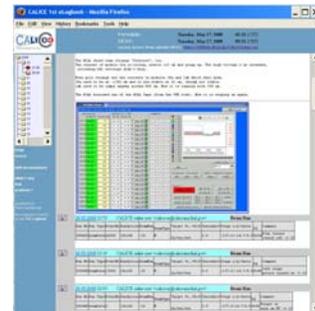

**Figure 6**: eLogbook

Since October 2001 the documentation of all accelerator operation relevant information at FLASH [3] is accomplished by using the "electronic logbook" developed by the DESY MCS4 group. Due to the broad acceptance and usage of this service this eLogbook [4] is also used at several other accelerator and detector facilities like CALICE.

Every standard web browser can function as a user interface for the input of text and retrieval of information. Also sending new entries by email is backed. Graphical data is inserted by low level postscript print services to offer a platform independent input interface. The generation of PDF is provided for high quality printouts. All data is stored in the today widely used XML format to allow high performance searches and interfacing with other web based services. A standard web server is generating dynamic content by use of JAVA servlet technology.